# Why are prices proportional to embodied energies?

Benjamin Leiva[a,*]

[a] CAES, University of Georgia, Athens, U.S.A.

[*] Corresponding author

*Email addresses*: bnlc@uga.edu (B. Leiva)

## Abstract

The observed proportionality between nominal prices and average embodied energies cannot be interpreted with conventional economic theory. A model is presented that places energy transfers as the focal point of scarcity based on the idea that (1) goods are material rearrangements, and (2) humans can only rearrange matter with energy transfers. Modified consumer and producer problems for an autarkic agent show that the opportunity cost of goods are given by their marginal energy transfers, which depend on subjective and objective factors (e.g. consumer preferences and direct energy transfers). Allowing for exchange and under perfect competition, nominal prices arise as social manifestations of goods' marginal energy transfers. The proportionality between nominal prices and average embodied energy follows given the relation between the latter and marginal energy transfers.

**Nomenclature**

| | | | | |
|---|---|---|---|---|
| $Q_{f,t}$ | Demand of final good f during $t$ | | $\varepsilon_l$ | $l$'s direct energy transfer |
| $Q_{e,t}$ | Demand of energy good $e$ | | $p_l$ | $l$'s power rate |
| $Q_{l,t}$ | Demand of prime mover (PM) $l$ | | $\phi_{l,t}$ | $l$'s power scarcity cost/energy surplus |
| $\tau_{k,t}^A$ | Av. energy transfer of any good $k$ | | $d_l$ | $l$'s depreciation rate |
| $\tau_{k,t}$ | Mg. energy transfer of good $k$ | | $g_{-l,k,t}^l$ | Mg. req. of production function |
| $\mu_k$ | Quantity elasticity (QE) of $\tau_{k,t}^A$ | | $\delta_e$ | Energy good $e$'s energy content |
| $\eta_k$ | QE of average embodied energy of $k$ | | $\Theta_t$ | Over-assignment of energy surplus |
| $E_t$ | Energy surplus | | $\Lambda_{f,t}$ | Energy assigned per unit of $f$ |
| $\lambda_t$ | Marginal utility of energy surplus | | $p^c$ | Commodity price |
| $\beta_{t,i}$ | $i$-period discount factor at $t$ | | $p_k$ | Real price of any good $k$ |
| $\psi_{k,t}$ | Direct energy transfers on $k$ | | $P_k$ | Nominal price of any good k |
| $\theta_{k,t}$ | Av. power scarcity cost of mg. $k$ | | $\tau_m$ | Mg. energy transfer of real money |
| $\Gamma_{k,t}$ | Embodied energy of PM producing $k$ | | $\tau_s$ | Syn. energy transf. of nom. money |
| $x_{l,k,t}$ | Quantity of $l$ producing $k$ | | $Q_m$ | Quantity of real money |
| $\tilde{x}_{l,t}$ | Endowment of $l$ | | $Q_n$ | Quantity of nominal money |

# 1. Introduction

Nominal prices and average embodied energies[1] seem to be directly proportional (Gutowski et al., 2013; Liu et al., 2008). Economic theory cannot explain this proportionality because if energy is like any other input (as conventional theory suggests), energy's cost share should be systematically high, yet estimates are consistently below 10% (Ayres et al., 2013; Csereklyei et al., 2016; Lindenberger & Kümmel, 2011).

Other interpretations are unavailable with neoclassical micro theory and all leading macro theories because energy is not part of their core constructs (Jehle & Reny, 2011; Mas-Colell et al., 1995; Samuelson & Nordhaus, 2004; Silberberg & Suen, 2001). Such omission is surprising considering the accepted truths that (1) goods are material rearrangements (Ryan & Pearce, 1985; von Mises, 1949), and (2) humans can only rearrange matter with energy transfers. Together, these ideas suggest that means are energy transfers, as supply depends on the factors that influence energy transfers: primarily the energy goods that provide energy (e.g. rice, oil) and prime movers that transfer it (e.g. workers, engines), and secondarily whatever other factors that influence energy transfers processes (e.g. water, information, social norms, the environment).

The only available alternative to economic theory that could interpret the proportionality between nominal prices and average embodied energies is an energy theory of value (Costanza, 1980; Hannon, 1973; Odum, 1971). Yet, such theory creates more problems than it solves. Imposing that economic value is defined by the energy spent producing goods severely reduces or ignores the role of intra and inter-temporal preferences, other inputs, and technological progress (Alessio, 1981; Hertzmark, 1981; Huettner, 1982; Huettner, 1976; Webb & Pearce, 1975). Furthermore, energy's relevance cannot be understood independently from the prime movers that transfer it to rearrange matter, e.g. a barrel of oil is useless without an engine that can transfer its energy content to produce a good.

Hence, why are nominal prices proportional to average embodied energies? This paper attempts to theoretically explain such proportionality following conventional economic rationale. The starting point is an autarkic agent that maximizes utility subject to an energy *transfer* constraint. Such constraint, justified with the idea that means are energy transfers, is the point of departure from conventional theory. Given this constraint, the paper shows how maximizing consumer and producer behavior reveals marginal energy transfers, what influences their magnitudes, and that nominal prices arise as their social representation under exchange and perfect competition. The proportionality under study follows given the relation between marginal energy transfers and average embodied energies.

---

[1] Average embodied energy is the total energy required on average to produce a unit of a good.

The remainder of the paper is as follows: Section 2 lays out a model in which an autarkic agent maximizes utility subject to an energy transfer constraint, and given such constraint, solves an array of optimization problems that must be solved for Pareto-efficiency. Section 3 extends the autarkic setting to analyze how exchange leads to nominal prices as representations of marginal energy transfers, and relates such magnitudes to average embodied energies. Section 4 provides a discussion on these results and section 5 concludes.

## 2. A revised theory of choice

My point of departure from traditional theory is the consideration of means as energy transfers. Given this proposition, utility maximization for an autarkic agent is subject to an energy transfer constraint, and given such constraint, an array of secondary optimization problems must be solved for Pareto-efficiency. This section discusses why means are energy transfers, and solves both primary and secondary optimization problems.

Means are energy transfers because humans can only act by rearranging matter, and they can only rearrange matter with energy transfers. Even if ends are immaterial such as reputation, power, or love, they can only be obtained with material rearrangements such as the writing of a seminal paper, the conquest of a territory, and the shipment of flowers. As Adams (1982) puts it *"Every event in history can occur only insofar as there is available whatever amount of energy (i.e., work) is necessary to carry it out. We can think thoughts wildly, but if we do not have the wherewithal to convert them into action, they will remain thoughts"*.

The proposition that means are energy transfers requires two clarifications. One is that it is teleological, as energy transfers are means only insofar as they do material rearrangements in accordance to human ends. Such rearrangements are economic goods (goods hereafter), which excludes free goods untouched by human action such as freely available clean air, and material rearrangements that do not serve human purpose such as an AC unit in the arctic.

The second clarification is that energy transfers are a process, not an object. The inputs enabling energy transfers can be categorized as energy goods providing energy (e.g. rice, oil), prime movers transferring such energy (e.g. laborers, engines), raw material being rearranged (e.g. wood, copper), and supporting inputs that enable such rearrangements (e.g. water, infrastructure). Thus, humanity can be constrained by its ability to transfer energy even when more energy reaches earth from the sun in an hour than what humanity currently uses in a year. Energy transfers are mostly limited by prime mover availability and the incapacity to transfer sunlight into energy goods at an energy surplus, but in general they can be constrained by the scarcity of any input.

The constraining role of scarce inputs do not contradict the fact that energy transfers are the focal point of scarcity. Inputs are scarce only in relation to the amount needed to do such transfers and can be substituted according to their relative convenience. On the contrary, there are no substitutes for energy transfers as the way humans alter their surroundings, and therefore outputs are scarce only in relation to the amount of energy transfers they require. In a hypothetical world where more energy transfers could be done than those required for bringing about and maintaining Eden, the economic problem would cease to exist. This is why human progress is associated with the capacity to transfer energy (White, 1943; Cook, 1976; Weissenbacher, 2009). Early societies could only transfer the energy content of some types of biomass with the human body, while modern ones complement such transfers with the energy content of oil, uranium, electricity etc. transferred by diesel engines, steam turbines, computers, and a host of other prime movers.

### 2.1 Utility maximization

Consider a multi-period autarkic agent that faces no uncertainty and has preferences represented by a quasiconcave, continuous and twice-differentiable utility function of the form

$$U = U\big(\boldsymbol{Q}_{F,1}, \boldsymbol{Q}_{F,2}, \dots, \boldsymbol{Q}_{F,T}\big), \tag{1}$$

where $T$ is the agent's planning horizon and $\boldsymbol{Q}_{F,t} = [Q_{1,t}, \dots, Q_{F,t}]'$ are final goods demanded in period $t$. Energy goods and prime movers are excluded from the utility function as a simplifying assumption and all goods are assumed perfectly divisible. The agent's primary objective is to maximize $U$ subject to each period's energy transfer constraint

$$\tau_{F,t}^{A\prime} \boldsymbol{Q}_{F,t} \leq E_t, \qquad \forall\, t \in T \qquad (2)$$

where $\tau_{F,t}^{A\prime} = [\tau_{1,t}^{A}, \dots, \tau_{F,t}^{A}]'$ are final goods' *average energy transfers* representing the average opportunity cost of producing each unit,[2] and $E_t$ is the energy surplus secured in $t$. This constraint is real because it is a physical imperative that the energy transferred in the production of final goods (the left-hand side) does not exceed the energy surplus from where energy is transferred (the right-hand side). Moreover, the constraint is reasonable despite being unobservable and arguable unknown because this is an autarkic agent for which an observable monetary constraint does not exist. The relevant Lagrangian at $t = 1$ is

$$\mathcal{L}_f = U(\boldsymbol{Q}_{F,1}, \boldsymbol{Q}_{F,2}, \dots, \boldsymbol{Q}_{F,T}) + \sum_{t=1}^{T} \lambda_t \left( E_t - \tau_{F,t}^{A\prime} \boldsymbol{Q}_{F,t} \right), \qquad (3)$$

where $\lambda_t$ is the marginal utility of energy surplus in period $t$. The FOCs to choose optimal quantity of final good $f$ are

$$U_{f,t}/\lambda_t = \tau_{f,t}, \qquad \forall\, f \in F, t \qquad (4)$$

where $\tau_{f,t} = \tau_{f,t}^{A}\left(1 + \mu_f\right)$ is good $f$'s *marginal* energy transfer, i.e. the minimum energy transfer required to produce one more unit of the good, and $\mu_f = \frac{\partial \tau_f^{A}}{\partial Q_f} \frac{Q_f}{\tau_f^{A}}$ is the quantity elasticity of average energy transfer. Given that average and marginal energy transfers are exogenous for the agent *as a consumer*, the FOCs show that the agent adjusts consumption levels such that in equilibrium the MRS between a final good and energy surplus equals the good's marginal energy transfer. Thus, marginal energy transfers reflect ratios between final goods' contribution to the agent's ends ($U_{f,t}$), and the marginal utility of its energy surplus ($\lambda_t$). The latter represents the general intensity of the agent's desires: if $\lambda_t$ is high then desires are strong, and an additional Joule of energy can substantially increase utility.

The agent's intertemporal preferences are defined by the time-shape of discount factors, where the $i$-period discount factor at $t$ is endogenously defined as

$$\beta_{t,i} = \lambda_{t+i}/\lambda_t, \qquad \forall\, t, i \qquad (5)$$

---

[2] A formal definition of energy transfers is provided in section 3.2. The problems of system boundary and co-products are overlooked by assuming that all energy transfers can be uniquely allocated among goods.

and represents the MRS between energy surplus during $t + i$ and $t$. Given this definition of $\beta_{t,i}$, ratios of (4) shows that optimal demand for good $f$ $i$-periods apart must observe

$$U_{f,t+i}/U_{f,t} = \beta_{t,i}\,\tau_{f,t+i}/\tau_{f,t}, \qquad \forall\, t, i \qquad (6)$$

which is consistent with Euler's equation despite the endogenous discount factor. $\beta_{t,i}$ could be greater than unity (Hansen & Singleton, 1983; Hotz et al., 1988; Hurd, 1989) because the relative magnitudes of $\lambda_t$ depend on the relative magnitudes of $E_t$, or in other words, because an agent's time preference depends on the time-shape of his income stream (Fisher, 1930). If $\beta_{t,i} \leq 1\ \forall\, i$, which is consistent with the growth of energy surplus over time, then $\beta_{t,i}$ is non-increasing in $i$, and typically decreasing. Yet, $\beta_{t,i}$ can increase (or decrease) over $t$ while decreasing over $i$.

## 2.2 Marginal energy transfers

Minimizing direct energy transfers in the production of any good $k \in K$ is a Pareto-efficient condition to maximize utility because it is required to minimize the left-hand side of the energy budget constraint in (2). The resolution of this problem, detailed in part A of the appendix, shows that good $k$'s marginal energy transfer is

$$\tau_{k,t} = \psi_{k,t} + \theta_{k,t}, \qquad \forall\, k \in K, t \qquad (7)$$

where $\psi_{k,t} = \frac{1}{L}\sum_{l=1}^{L}\varepsilon_l g'_{-l,k,t}$ is the average direct energy transfers required to produce one more unit of good $k$, and $\theta_{k,t} = \frac{1}{L}\sum_{l=1}^{L}\phi_{l,t} g'_{-l,k,t}$ is the average power scarcity cost in the production of an additional unit of good $k$. Power scarcity costs are due to limited prime movers, and thus marginal energy transfers account for both energy goods and prime movers. Moreover, $L$ is the quantity of prime mover types used by the agent, and for any type $l$, $\varepsilon_l = \int_t^{t+1} p_l\, dt$ is the direct energy transfer, $p_l$ is the effective power rate, $g'_{-l,k,t} = f_{l,k}^{-1}$ is the marginal requirements of production function (with $f_{l,k}$ the marginal productivity in the production of good $k$), and $\phi_{l,t}$ is the power scarcity cost. A prime mover's power scarcity cost derives from its limited availability, and is measured in Joules as it reflects the energy surplus that could be secured with an additional unit of a prime mover (see part B of the appendix).

Equation (7) shows that marginal energy transfers are endogenous for the agent *as a producer*, and depends on prime movers' power rate, productivity, and scarcity. Marginal energy transfer change upon changes in $\varepsilon_l$ due to variations in effective power rates, in $g'_{-l,k,t}$ due to changes in management or the environment, and in $\phi_{l,t}$ due to shifts in demand for $l$ in the production of any good. In a world with uncertainty, new prime movers and energy goods introduced in a period $t_1$ would shock the values of $\tau_{k,t \geq t_1}$ by modifying $g'_{-l,k,t \geq t_1}$ and $\phi_{l,t \geq t_1}$ respectively.

### 2.3 Equilibrium magnitudes

Marginal energy transfers are defined by (7), but in equilibrium they equal expressions that vary according to the nature of good $k$. These expressions are derived from three distinct optimization procedures. Their resolution, alongside energy transfer minimization, is a Pareto-efficient condition for utility maximization. The first optimization procedure is energy surplus maximization, which is required to maximize the right-hand side of the energy transfer constraint in (2). The resolution of this problem (part B of the appendix) shows that in equilibrium the marginal energy transfer of energy good $e \in \epsilon \in K$ is

$$\tau_{e,t} = \beta_{t,1} \delta_e, \qquad \forall\, e \in \epsilon, t \qquad (8)$$

where $\beta_{t,1}$ is the one-period ahead discount factor during $t$ and $\delta_e$ is good $e$'s energy content in Joules.[3] Equation (8) implies that energy goods' marginal energy transfers are influenced by their energy content and the agent's intertemporal preferences, as in equilibrium the agent equates energy goods' opportunity cost to the discounted energy income they provide.

The second procedure is optimal prime mover accumulation, which is required to minimize the left-hand side of the energy transfer constraint in (2). The resolution of this problem, (part C of the appendix) shows that in equilibrium the marginal energy transfer of prime mover $l$ is

$$\tau_{l,t} = \sum_{i=1}^{T} \beta_{t,i} \phi_{l,i+1} d_l^{i-1}, \qquad \forall\, l \in L, t \qquad (9)$$

where $\beta_{t,i}$ is the $i$th-period ahead discount factor during $t$ and $d_l \in (0,1)$ is $l$'s linear depreciation rate. Given that $\phi_{l,t}$ can be interpreted as prime movers' marginal energy surplus (see part B of the appendix), equation (9) implies that at equilibrium, the opportunity cost of prime movers —i.e. their marginal energy transfer— equals the future flow of discounted net energy surplus that they generate. By extension, the opportunity cost of all assets is the present value of the flow of energy surplus that they generate, or as Fisher (1930) puts it "*The value of any property is its value as a source of income and is found by discounting that expected*

---

[3] We overlook other features of energy goods, such as density and cleanness. For a discussion on energy goods see Cleveland et al., (2000), Podobnik (2005), Stern (2010), Bhattacharyya (2011), and Smil (2016).

*income*". Furthermore, although prime movers produce the energy goods that build energy surplus, with the relation going from the former to the latter, the value of such prime movers is determined by the energy surplus they secure, with the relation going from the latter to the former. In short, "*Income values produce capital values*" (Fisher, 1930).

The third and last optimization procedure is optimal allocation of energy surplus, which is required for the quantities produced and demanded of final goods to be equal. The resolution of this problem, (part D of the appendix) shows that in equilibrium the marginal energy transfer of final good $f$ is

$$\tau_{f,t} = (1 - \Theta_t)\Lambda_{f,t}, \qquad \forall f, t \qquad (10)$$

where $\Theta_t \in [0,1)$ is a measure of the agent's over-assignment of energy surplus during $t$, and $\Lambda_{f,t}$ is the energy assigned by the agent to the production of each unit of final good $f$. Equation (10) implies that final goods' marginal energy transfer is influenced by their energy assignments and excess of energy assignments.

Using equations (4) and (10) shows that effective energy assignments are

$$(1 - \Theta_t)\Lambda_{f,t} = U_{f,t}/\lambda_t, \qquad \forall f, t \qquad (11)$$

where the right-hand side is the MRS between a final good and energy surplus. Thus, the agent chooses consumption levels *as a consumer* such that the endogenous marginal utility of final goods leads to a MRS with respect to energy surplus equal to the good's exogenous marginal energy transfer. Also, the agent chooses production levels *as a producer* such that the endogenous marginal energy transfer of final goods equal the goods' exogenous effective energy assignments. Equilibrium is reached when effective energy assignments equal the MRS between goods and energy surplus.

## 3. A revised theory of price

Under autarky no nominal prices exist. This section allows exchange to take place between previously autarkic agents, and shows how nominal prices arise through exchange and competition as social representations of marginal energy transfers. Furthermore, the section formally relates nominal prices to average embodied energies.

Agents have incentives to engage in exchange because, in general, doing so further increases their utility. Exchange leads to gains from trade from the lowering of marginal energy transfers, which relaxes their energy transfer constraints. These gains follow conventional Ricardian logic.

### 3.1 Commodity prices

Assume two previously autarkic agents exchange two goods and consume them in the same quantities as under autarky. Also assume exchange takes place in a single period such that the time subscript can be omitted. If "agent 1" exchanges $q_B$ units of good B for $q_C$ units of good C with "agent 2", the gains from trade ($GT$) are

$$GT = \int_{q_B^E}^{q_B^A} (\tau_B^2 - \tau_B^1) dq_B + \int_{q_C^E}^{q_C^A} (\tau_C^1 - \tau_C^2) dq_C - TC \tag{12}$$

where $q_i^A$ and $q_i^E$ are the quantities produced of good $i = B, C$ under autarky and exchange respectively, and where $q_B = q_B^A - q_B^E$ and $q_C = q_C^A - q_C^E$. The positive terms are savings from reduced production; the negative ones are spending from increased production; and $TC$ are transaction costs due to transportation, spoilage, etc. When $GT > 0$ there are unambiguous gains from trade because production and consumption of both goods is the same under exchange as under autarky, but at lower average energy transfers. Gains from trade are a "release" of energy surplus that the agents can use to produce other goods. If under autarky $\tau_B^1 < \tau_B^2$ and $\tau_C^1 > \tau_C^2$ ($\tau_B^1 < \tau_B^2$ and $\tau_C^1 < \tau_C^2$) there are absolute (relative) comparative advantages.

For the unambiguous gains from trade to benefit both agents, the rate of exchange of goods $q_B/q_C$ (i.e. their commodity price) must remain between agents' reservation values. Such values are $\tau_C^1/\tau_B^1$ for agent 1 and $\tau_C^2/\tau_B^2$ for agent 2. The commodity price that will govern this exchange is between agents' reservation values, yet nothing more is known as both agents are monopolies of the good they produce, and monopsonies of the other one. The exact commodity price will depend on site-specific features such as negotiating abilities.

Changes in marginal and average energy transfers alter each agent's entire schedule of production and consumption. As shown in Figure 1, the intersection between a good's demand and marginal energy transfer curve (METC) yields equilibrium production and consumption under autarky. The agent in the left diagram has a relatively lower METC, and under exchange will produce more relative to autarky. Faced with a higher marginal energy transfer $\tau^{E'} > \tau^E$, this agent will demand less, and the difference will be exported (E). The agent in the right diagram, faced with a lower marginal energy transfer $\tau^{I'} < \tau^{I}$, will produce less and demand more. The difference are imports (I). Under exchange, the marginal energy transfer of both agents is the same $\tau^{E'} = \tau^{I'}$.

Without the assumption that the agents' consumption remains the same after exchange there are still unambiguous gains for both agents if $GT > 0$ and if the rate of exchange of goods is between agents' reservation values. If an agent willingly reduces the consumption of a good as compared to autarky, doing so must provide the agent with another good that yields more utility.

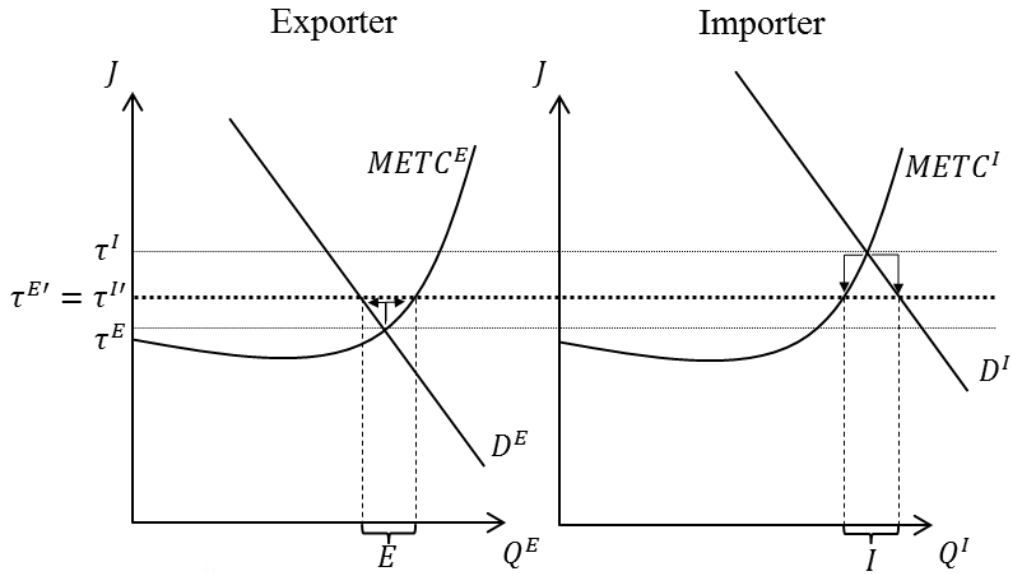

Figure 1: Changes in demand and supply due to exchange.

The optimal rate of exchange of goods is given by maximizing (12) choosing $q_B$ and $q_C$. With either good the FOC is

$$\tau_i^I = \tau_i^E + MTC_i \qquad (13)$$

where $\tau_i^I$ is the marginal energy transfer of good $i = B, C$ of the importing agent $I = 1,2$, $\tau_i^E$ is the marginal energy transfer of good $i$ of the exporting agent $E = 1,2$, and $MTC_i$ is the marginal transaction cost of good $i$. This condition sets the patterns of trade: until (13) is observed, agents with relatively lower marginal energy transfers will export a good, while those with relatively higher ones will import it. If $TC = 0$ (assumed hereafter), the optimal quantity exchanged by the agents equalizes the marginal energy transfer of each good between agents (Figure 1).

Under (11) the agents exchange $q_B^*$ for $q_C^*$, and the prevailing marginal energy transfers are $\tau_B$ and $\tau_C$. Marginal energy transfers are the same for both agents given $MTC_i = 0$. Thus, optimal exchange implies the commodity price

$$q_B/q_C = p^c = \tau_C/\tau_B. \tag{14}$$

The condition in (13) and the price associated with it in (14) materializes through competition. Adding more agents to the exchange of goods A and B makes negotiation abilities and other site-specific features progressively irrelevant. As monopolistic and monopsonistic positions are lost, Nash equilibrium brought about by tâtonnement leads to the collapse of the range of possible commodity prices at the ratio of marginal energy transfers under exchange $\tau_C/\tau_A$.

### 3.2 Real and nominal prices

Even a modest number of $K$ goods make barter inviable. Agents convene on using real money (i.e. commodity money or general equivalent) to solve this problem (Menger, 1892; Schumpeter, 1954). If compliant with certain characteristics (Einzig, 1966), real money solves the problem of double coincidence of wants as a medium of exchange, facilitates savings as a store of value, enables debt as a unit of deferred payments, and reduces the number of commodity prices from $K!$ to $K - 1$ as a unit of account ( Karimzadi, 2013; Schumpeter, 2014; Smithin, 2014).

Despite acquiring such attributes, real money is first and foremost a good such as cowries, cows, or gold (Cribb, 1986), and therefore has its own marginal energy transfer $\tau_m$. Given (14), the commodity price of commodity $k$ in terms of real money, i.e. its real price, is

$$p_k = \tau_k/\tau_m, \tag{15}$$

which can have complex movements as both $\tau_k$ and $\tau_m$ can be simultaneously changing. For example, $\tau_m$ can increase due to depletion of gold deposits, and $\tau_k$ can fall due to innovations reducing energy transfers or investments reducing prime mover constraints.

Given the hurdles inherent in using real money (Karimzadi, 2013), societies further facilitate exchange by adopting nominal money as a representation of real money (Poor, 1969). Nominal money, also called representative money or currency, does not derive its rate of exchange with other goods from its own marginal energy transfer. Nominal money such as pesos, dollars, and euros have a "synthetic energy transfer" established by applying (14) between real and nominal money. The synthetic energy transfer of nominal money, i.e. its purchasing power, is

$$\tau_s = \tau_m Q_m / Q_n \qquad (16)$$

where $Q_m$ and $Q_n$ are respectively the total quantities of real and nominal money in an economy. Inverting, logging, and differentiating (16) shows the rate of decrease of the purchasing power of nominal money, i.e. inflation, as

$$\pi = dln(Q_n/Q_m) - d\ln(\tau_m), \qquad (17)$$

where $\pi = dln(1/\tau_s)$. Given (14), the real price of commodity $k$ in terms of nominal money, i.e. its nominal price, is

$$P_k = \tau_k / \tau_s, \qquad (18)$$

which holds for all goods regardless if they are final, energetic, or prime movers. Thus, when agents engage in exchange using nominal money and prices, it is unknown and irrelevant to them that marginal energy transfers are defined according to (7) and expressed according to (18). The same is true regarding how marginal energy transfers of energy goods, prime movers, and final goods are influenced according to (8), (9), and (10) respectively. Under the cloak of nominal prices all goods appear equal.

The movements of nominal prices also contribute to such mimetization. Logging and differentiating (18), and replacing using (16) shows the rate of change of nominal prices as

$$dln(P_k) = d\ln(p_k) + dln(Q_n/Q_m), \qquad (19)$$

which implies that all nominal prices are affected by the same common nominal component $dln(Q_n/Q_m)$. Equation (19) shows that nominal prices experience more complex movements than real prices, as the former contains the latter and is additionally influenced by the rate of growth of the ratio of nominal money to real money. In fact, $dln(Q_n/Q_m)$ can lead to drastic changes in nominal prices (e.g. expansive monetary policies or trade deficits under the gold standard) even when changes in the real component is zero, i.e. $dln(p_k) = 0$. Logically, if most of these variables remain stable, then nominal and real prices can be mostly stable. Thus, equation (19) interprets the existence of waves of price equilibriums and price revolutions as documented in Fischer (1996), although such interpretation exceeds the scope of this article.

Caution must be taken when nominal money is replaced with fiat money. Then, $\tau_s$ stops being defined by (16) and thus the following equations become invalid. In such context, $\tau_s$ starts depending on the trust that agents have on the issuer of fiat money. The debate between Chartalism (Wray, 2012) and Metallism (Menger, 1892) exceeds the scope of this paper.

The ratios of real prices offset the variability in real prices that stem from the marginal energy transfer of real money. Considering commodities $A$ and $B \in K$ with real prices $p_A = \tau_A/\tau_m$ and $p_B = \tau_B/\tau_m$ shows that their relative real price equals the ratio of their marginal energy transfers

$$p_A/p_B = \frac{\tau_A/\tau_m}{\tau_B/\tau_m} = \tau_A/\tau_B. \tag{20}$$

Similarly, relative nominal prices offset the variability in nominal prices that stem from the purchasing power of nominal money, and they also offset the effect of fiat money. Considering nominal prices $P_A = \tau_A/\tau_s$ and $P_B = \tau_B/\tau_s$ shows that their relative nominal price also equals the ratio of their marginal energy transfers

$$P_A/P_B = \frac{\tau_A/\tau_s}{\tau_B/\tau_s} = \tau_A/\tau_B. \tag{21}$$

### 3.3 The proportionality between prices and embodied energies

Equations (16) and (19) do not directly interpret the proportionality under study because they contain marginal energy transfers instead of average embodied energies. Yet, they come close as both concepts are tied by the direct energy transfers required to produce goods. Good $k$'s marginal embodied energy can be decomposed as

$$\gamma_k = \psi_k + \Gamma_k, \tag{22}$$

where $\psi_k$ is good $k$'s marginal direct energy transfer as described in (7) and $\Gamma_k$ is the average energy transfers used to build the prime movers producing the marginal unit of $k$. Thus, marginal energy transfers and embodied energies coincide regarding direct energy transfers yet differ regarding indirect transfers. Embodied energies are backward-looking by considering as indirect transfers the energy used in the past to produce the prime movers doing direct transfers (Chapman, 1974; IFIAS, 1974). Energy transfers are forward-looking by considering as indirect transfers the power scarcity of prime movers doing direct transfers. As this measure is determined by prime movers' potential to provide energy surplus in the future and have no relation to the past, energy transfers recognize that "bygones are forever bygones".

Given (7) and (22), marginal energy transfers and embodied energies are related as

$$\tau_k = \gamma_k + (\theta_k - \Gamma_k), \tag{23}$$

such that they differ according to the difference between prime movers' potential to provide energy surplus and the energy transferred to produce them. Another reason equations (18) and (21) fail to interpret the proportionality under study is because they contain marginal values, while estimates of embodied energies are of average embodied energies or of random samples of the marginal embodied energy curve. As in either case the most reasonable assumption is that available estimates are averages, then replacing (18) with (23) and noting that $\gamma_k = \gamma_k^A(1 + \eta_k)$ leads to nominal prices as

$$P_k = \frac{\gamma_k^A(1+\eta_k) + (\theta_k - \Gamma_k)}{\tau_S}, \tag{24}$$

where $\eta_k = \frac{\partial \gamma_k^A}{\partial Q_k} \frac{Q_k}{\gamma_k^A}$ is good $k$'s industry-wide quantity elasticity of average embodied energy. Equation (24) interprets the proportionality under study by indicating a noisy positive relation between nominal prices and average embodied energies. The noise is due to differences between the energy transferred to produce prime movers and their potential to provide energy surplus, differences between marginal and average values, and inflation.

To control for the effect of inflation, the ratio of (24) for any two commodities $A$ and $B \in K$ is

$$\frac{P_A}{P_B} = \frac{\gamma_A^A(1+\eta_A) + (\theta_A - \Gamma_A)}{\gamma_B^A(1+\eta_B) + (\theta_B - \Gamma_B)}. \tag{25}$$

This equation shows that, *ceteris paribus*, goods with relatively lower returns to scale (higher $\eta_k$) will command a relatively higher nominal price than what their relative average embodied energy alone would suggest.

## 4. Discussion

The proportionality between market prices and average embodied energies is a consequence of recognizing energy transfers as the means available to human ends. Nominal prices represent marginal energy transfers, and such transfers are related to average embodied energies. The proportionality under study is noisy due to the differences between energy transfers and embodied energies, the difference between marginal and average embodied energy, and inflation.

The model presented is mostly consistent with conventional economic theory. Economics is the study of human behavior as a relationship between ends and scarce means (Robbins, 1932); means are simply energy transfers. Production is not creation but transformation through arrangement (Ryan & Pearce, 1985; von Mises, 1949); arrangements are simply argued to be a material process associated with energy transfers. Production consists of the rendering of services by human agencies (Friedman, 1962; Knight, 1935); services are simply specified as energy transfers and human agencies as prime movers.[4] The cost of a service in producing a good is the other goods it could produce (Stigler, 1987); the Joule only measures the quantity of goods that are being sacrificed. In summary, we can recognize means as energy transfers and agree with all fundamental laws of neoclassical microeconomics (Rappaport, 1998).

Disagreements between conventional theory and this model stem from energy's place in economics. While energy is ignored by the former, the latter recognizes that energy transfers are the means by which humans act. This leads to several divergences, the two most important being the understanding of the limiting factors of production and growth, and the acceptance of a physical unit of value. These novelties signal where this perspective provides new insights into economic phenomena. The first one suggests that growth is constraint by whatever hinders the capacity to transfer more energy, and therefore highlights the importance of prime mover accumulation, efficiency enhancements, and energy goods' energy surplus. The second suggests that by measuring the opportunity cost of goods in Joules, market prices can be interpreted as social manifestations of underlying energy magnitudes. Although controversial for economists, this proposition simply observes that "*Energy is the only universal currency: one of its many forms must be transformed to another in order for stars to shine, planets to rotate, plants to grow, and civilizations to evolve*" (Smil, 1999).

This perspective does not support an energy theory of value or an energy transfer theory of value. First, means as energy transfers do not imply valuations that are independent from human ends. Material rearrangements become goods and energy transfers means only if they further human ends. Furthermore, marginal energy transfers are goods' opportunity cost, and therefore a comparison that respects the chief notion that nothing can be valuable without reference to something else. Whatever marginal energy transfers are revealed in a given economy, they are specific to the ends sought by the unique human beings constituting it, and on the factors influencing the scarcity of energy transfers in a given place and time. Second, although nominal prices represent marginal energy transfers, such magnitudes depend on an array of variables apart from direct energy transfers and the scarcity of prime movers. Marginal energy transfers of final goods also depend on the MRS between goods and energy surplus, and those of energy goods and prime movers on the agent's time preferences. If anything, this perspective harmonizes objective and subjective schools of economic value by recognizing human ends as the driver of economic activity, and energy transfers as the means to fulfill them.

---

[4] Making such specification in Knight (1935), page nine, second paragraph, is highly illuminating.

Further clarifications are in order on why this perspective does not support an energy theory of value, and how it responds to the issues put forth in Alessio (1981), Hertzmark (1981), Huettner (1976, 1981) and others. First, this perspective highlights energy transfers and not energy as the ultimately scarce resource, and notes that energy transfer is a process done by prime movers using energy goods while supported by a host of other factors such as water, infrastructure, social norms, etc. Second, there is nothing in the model requiring the scarcity of energy goods to intensify over time. This is likely to happen, and is required for a long run equilibrium to exist, but a society that manages to produce ever more energy goods at constant marginal energy transfers will simply grow indefinitely. Third, there is no suggestion that a pure physical energy analysis is superior to a monetary analysis. On the contrary, monetary magnitudes are expressions of underlying energy flows, and thus conventional economic analysis is energy analysis in disguise. Fourth, price levels and relative prices are not dependent on supply alone, as marginal energy transfers depend on intra and inter-temporal preferences. Fifth, time preferences heavily weight on the marginal energy transfers of energy goods and prime movers. Sixth, marginal energy transfers take into consideration all inputs required to produce a good, not only the energy goods and prime movers used in production. If an input (e.g. water) becomes scarce in some place and time, this means that more energy transfers are required to obtain it, which affects the marginal energy transfers of all goods using it. Seventh, technology plays a major role influencing marginal energy transfers and the magnitude of energy surplus. The disruptions of discoveries and inventions such as electricity and the induction motor are not analyzed because uncertainty was ruled out, not because they cannot fit within the model.

The model provides several secondary results. Energy transfers as a forward-looking measure of energy flow contrasts with the conventional backward-looking concept of embodied energy, and thus has stronger economic intuition. Similarly, a prime mover's marginal energy transfer (and nominal price) equals the discounted future energy surplus it will generate. Moreover, an endogenous discount factor and its relation to the agent's mEROI provides a new perspective on interest rates as the bridge between a society's impatience and opportunity. Lastly, the logical sequence leading to the use of real and nominal money provides a definition of inflation that is partially different to its conventional understanding.

Recognizing means as energy transfers has consequences that exceed the interpretation of the proportionality under study. Viewing nominal prices as representations of marginal energy transfers adds a new layer to microeconomic analysis. This is irrelevant if nominal prices are given, yet has implications when they are unreliable or nonexistent as with market power, price formation, and nonmarket valuation. Furthermore, this layer provides a new micro-foundation for macroeconomics, with implications for the understanding of growth, interest rates, inflation, and inequality. Proper accounts of these implications are future research avenues.

The model's limitations provide other topics for further research. Excluding energy goods and prime movers from agents' utility functions implies a relatively low loss of generality, yet their inclusion would yield a more general framework. More generality could also be obtained with the inclusion of institutional and environmental factors that influence economies and markets. Although such factors only stimulate or obstruct the fundamental tendencies that the current model describes, a complete description of nominal prices should consider them. Finally, the model is set up without uncertainty, market power, or other prominent features of real world economies. Given the use of conventional economic rationale, extensions considering choice under uncertainty, game theory, and monopolies should follow seemly.

## 5. Conclusion

Nominal prices are proportional to average embodied energies because marginal energy transfers measure goods' opportunity costs, and because energy transfers and embodied energies are related concepts of energy flows. These results derive from recognizing energy transfers as the means available to human purpose, the relevance of an autarkic agent's energy transfer constraint, and the role that prime movers play in production. These deviations from standard theory are the basis to interpret the proportionality under study, and to include energy within neoclassical micro theory. Understanding nominal prices as social representations of underlying marginal energy transfers provides a new layer to microeconomics and a new micro-foundation to macroeconomics, with implications for interest rates, inflation, growth, and among others, inequality. At a broader level, such understanding suggests a perspective of economics as an interplay between human desires and thermodynamic processes.

## Appendix

### Part A. Energy transfer minimization

This secondary objective for the agent takes the form

$$\min_{\boldsymbol{x}_{k,t}} \boldsymbol{\varepsilon}' \boldsymbol{x}_{k,t}, \qquad \forall\, k, t \qquad (A.1)$$

subject to the productive targets

$$f_{k,t}(\boldsymbol{x}_{k,t}) = \bar{Q}_{k,t}, \qquad \forall\, k, t \qquad (A.2)$$

and the prime mover constraints

$$\sum_{k=1}^{K} x_{l,k,t} \le \tilde{x}_{l,t}, \qquad \forall\, l, t \qquad (A.3)$$

where $\boldsymbol{\varepsilon}' \boldsymbol{x}_{k,t}$ is the direct energy transferred to produce $Q_{k,t}$, $\bar{Q}_{k,t}$ is the target of production of good $k$ during $t$, and $f_{k,t}(\cdot)$ is a concave from above, continuous, and twice-differentiable production function. Also, $\boldsymbol{x}_{k,t} = [x_{1,k,t}, \dots, x_{L,k,t}]'$ is the set of prime movers used to achieve such production and $\tilde{x}_{l,t}$ is the agent's availability of prime mover $l$ during $t$.

The objective and production functions only include prime movers because — in an autarkic setting — prime movers contain all other inputs. Supporting inputs (e.g. desks, warehouses, transmission lines) are subsumed by a given prime mover endowment because $\tilde{x}_l$ includes everything required to transfer energy. Energy inputs (e.g. rice, gasoline, electricity) are accounted for by prime mover's $\boldsymbol{\varepsilon}$ as prime mover's energy transfer is a transferal of energy goods' energy content. Furthermore, there are no intermediate inputs because the agent runs the entire productive chains, and raw materials need not be considered as they are spontaneous material arrangements that the agent finds in nature. Finally, the opportunity cost of prime movers themselves is factored in through prime mover constraints as shown below.

The minimization of $\boldsymbol{\varepsilon}' \boldsymbol{x}_{k,t}$ given $\bar{Q}_{k,t}$ and $\tilde{x}_{l,t} \,\forall\, l, t$ is done by modifying conventional cost minimization procedures described in Silberberg & Suen (2001) or similar. The relevant Lagrangian at $t = 1$ is

$$\mathcal{L}_k = \sum_{t=1}^{T} \sum_{k=1}^{K} \lambda_t \boldsymbol{\varepsilon}' \boldsymbol{x}_{k,t} + \sum_{t=1}^{T} \sum_{k=1}^{K} \lambda_t \tau_{k,t} \big[ \bar{Q}_{k.t} - f_{k,t}(\boldsymbol{x}_{k,t}) \big]$$

$$- \sum_{t=1}^{T} \sum_{l=1}^{L} \lambda_t \phi_{l,t} \big( \tilde{x}_{l,t} - \sum_{k=1}^{K} x_{l,k,t} \big), \qquad (A.4)$$

where each good's marginal energy transfer appears as the Lagrange multiplier associated with its productive target constraint because it is goods' opportunity cost. Each energy flow is multiplied by its corresponding marginal utility of energy because the primary objective of the agent is to maximize utility, not some energy flow *per se*. The FOC to choose the optimal quantity of prime mover $l$ to produce good $k$ in period $t > 0$ is

$$\tau_{k,t} f_{l,k,t} = \varepsilon_l + \phi_{l,t}, \qquad \forall\, k, l, t \qquad (A.5)$$

where $f_{l,k,t}$ is the marginal productivity of prime mover $l$ in the production of good $k$ in period $t$, and $\phi_{l,t}$ is its power scarcity cost in that period. Multiplying (A.5) by $g'_{-l,k,t}$ (the inverse of $f_{l,k,t}$) and averaging over $l$ yields the expression for marginal energy transfer in (7).

The resolution of the Lagrangian in (A.4) involves a system of $L^2 \cdot K^2 \cdot T^3$ equations and unknowns (the $\lambda_t$ are exogeneous for the agent *as a producer*). This system, even in its simpler versions (e.g. $L = 2$; $K = 2$; $T = 3$), has no closed form solution under conventional functional forms of the production function. Numerical approximations are required to obtain optimal prime mover demands, marginal energy transfers, and power scarcity costs.

## Part B. Energy surplus maximization
This secondary objective for the agent takes the form

$$max_{\boldsymbol{Q}_{e,t}} E_t = \delta_e Q_{e,t-1} - \boldsymbol{\varepsilon}' \boldsymbol{x}_{e,t}, \qquad \forall\, e, t \qquad (B.1)$$

subject to the prime mover constraints

$$\sum_{k=1}^{K} x_{l,k,t} \leq \tilde{x}_{l,t}, \qquad \forall\, l, t \qquad (B.2)$$

where $Q_{e,t-1}$ is the quantity of energy good $e$ produced in period $t-1$ and available in $t$ and $Q_{e,0}$ is the initial endowment of the good. The relevant Lagrangian at $t = 1$ is defined as

$$\mathcal{L}_e = \sum_{t=1}^{T} \lambda_t [\sum_{e=1}^{\epsilon} (\delta_e Q_{e,t-1} - \boldsymbol{\varepsilon}' \boldsymbol{x}_{e,t}) + \sum_{l=1}^{L} \phi_{l,t} (\tilde{x}_{l,t} - \sum_{k=1}^{K} x_{l,k,t})], \qquad (B.3)$$

where the multiplication of each energy flow by its corresponding marginal utility of energy responds to the idea that energy surplus maximization is a Pareto-efficient condition for utility maximization.

The FOC to choose the optimal quantity of energy good $e$ in period $t > 0$ indicates that energy goods' marginal energy transfers equals their discounted energy income

$$\tau_{e,t} = \beta_{t,1}\delta_e. \qquad\qquad \forall\, e \in \epsilon \qquad\qquad (B.4)$$

Given $\beta_{t,1} = \lambda_{t+1}/\lambda_t$, equation (B.4) implies that the utility the agent derives from producing an additional unit of an energy good in period $t+1$ ($\lambda_{t+1}\delta_e$) equals the utility forgone due to its production in $t$ ($\lambda_t\tau_{e,t}$). Also, the condition implies that all energy goods have the same marginal Energy Return Over Energy Investment ($mEROI_{e,t} = \delta_e/\tau_{e,t}$) $\forall\, e, t$, yet not necessarily the same average Energy Return Over Energy Investment ($aEROI_{e,t} = \delta_e/\tau_{e,t}^A$).[5] This definition of $mEROI$ implies that $\beta_{t,1} = 1/mEROI_{e,t}$ $\forall\, e$, and thus a relation between the agent's time preferences, interest rate, and opportunities to increase energy transfers in the future.

The resolution of the Lagrangian in (B.3) involves a system of $L \cdot \epsilon \cdot T^3$ equations and unknowns. The equations representing the FOC with respect to $\phi_{l,t}$ are identical to the ones in the previous section, and thus provide no additional information. This system can be solved by considering $\phi_{l,t}$ exogeneous and the optimization based on energy transfers instead of direct energy transfers and prime mover constraints. The system yields optimal production of energy goods and the agent's entire schedule of energy surplus.

Prime movers' power scarcity cost is their marginal energy surplus, which is found replacing $Q_{e,t} = f_{e,t}(\boldsymbol{x_{e,t}})$ in (B.3). Deriving such expression with respect to any prime mover $x_{l,e,t}$, and averaging over $e$ shows that prime mover $l$'s power scarcity cost is

$$\phi_{l,t} = \beta_{t,1}\frac{1}{\epsilon}\sum_{e=1}^{\epsilon}(\delta_e f_{l,e,t} - \varepsilon_l), \qquad\qquad \forall\, l \qquad\qquad (B.5)$$

where the right-hand side is $l$'s discounted marginal energy surplus secured on average across all energy goods. This surplus provides the incentive for the agent to accumulate prime movers.

### Part C. Optimal prime mover accumulation

This secondary objective for the agent takes the form

$$max_{\boldsymbol{Q_{l,t}}}\,\xi_L = \sum_{t=1}^{T}\sum_{l=1}^{L}[\phi_{l,t}(\tilde{x}_{l,t} - \sum_{k=1}^{K}x_{l,k,t}) - D_{l,t}], \qquad\qquad (C.1)$$

where $\xi_L$ is a quasi-energy surplus that indicates the energy balance between energy income and transfers associated with prime mover production. $\xi_L$ is a *quasi*-surplus because prime movers do not provide an energy income directly. Therefore, $\xi_L$ is a transferal of energy surplus from energy sectors to non-energy sectors. Equation (C.1) implies that given a schedule of $\phi_{l,t}$ and of prime mover use $(x_{l,k,t})$, the agent chooses the quantity of prime movers to be produced to maximize energy surplus. Production of prime movers appear in (C.1) because prime mover endowment is

$$\tilde{x}_{l,t} = \sum_{i=0}^{t-1} d_l^i Q_{l,t-1-i}, \qquad \forall\, l, t \qquad (C.2)$$

where $Q_{l,t-1-i}$ is the quantity of prime mover $l$ produced in period $t-1-i$, and $Q_{l,0}$ is its initial endowment. Replacing (C.2) in (C.1) and including $\lambda_t$ as done above for all energy flows implies that the relevant Lagrangian at $t = 1$ is

$$\mathcal{L}_l = \sum_{t=1}^{T} \lambda_t \sum_{l=1}^{L} [\phi_{l,t} (\sum_{i=0}^{t-1} d_l^i Q_{l,t-1-i} - \sum_{k=1}^{K} x_{l,k,t}) - D_{l,t}]. \qquad (C.3)$$

The FOC to choose the optimal quantity of prime mover $l$ in period $t > 0$ indicates that, in equilibrium, prime movers' marginal energy transfer equals the discounted net sum of the energy surplus they provide over their lifetime

$$\sum_{i=1}^{T} \beta_{t,i} \phi_{l,i+1} d_l^{i-1} = \tau_{l,t}. \qquad \forall\, l, t \qquad (C.4)$$

Given $\beta_{t,i} = \lambda_{t+i}/\lambda_t$, the condition equates the utility the agent derives in all future periods from producing an additional unit of a prime mover in $t$ ($\sum_{i=1}^{T} \lambda_{i+1} \phi_{l,i+1} d_l^{i-1}$) to the utility forgone due to its production in that period ($\lambda_1 \tau_{l,t}$). The condition also yields prime mover's optimal production in each period, prime movers endowment for the next one, and the agent's aggregate power rate $P_t = \sum_{l=1}^{L} p_l \tilde{x}_{l,t}$.

The resolution of the Lagrangian in (C.3) involves a system of $L \cdot T$ equations and unknowns. This system can be easily solved based on energy transfers instead of direct energy transfers and prime mover constraints. The system yields optimal production of prime movers and the agent's entire schedule of power.

## Part D. Allocation of energy surplus

This secondary objective for the agent takes the form

$$max_{Q_{f,t}} \xi_F = \Lambda_{f,t} Q_{f,t} - D_{f,t}, \qquad \forall\, f, t \qquad (D.1)$$

subject to the prime mover constraints

$$\sum_{k=1}^{K} x_{l,k,t} \leq \tilde{x}_{l,t}, \qquad \forall\, l, t \qquad \text{(D.2)}$$

and the energy assignment constraint

$$\sum_{f=1}^{F} \Lambda_{f,t} Q_{f,t} \leq E_t, \qquad \forall\, t \qquad \text{(D.3)}$$

where $\boldsymbol{\Lambda_t} = [\Lambda_{1,t}, \dots, \Lambda_{F,t}]'$ is the unitary energy assigned by the agent to the production of final goods. The last constraint binds the total energy assigned in the production of all final goods to the agent's energy surplus. The relevant Lagrangian at $t = 1$ is

$$\mathcal{L}_f = \sum_{t=1}^{T} \sum_{f=1}^{F} \lambda_t \left( \Lambda_f Q_{f,t} - D_{f,t} \right) + \sum_{t=1}^{T} \sum_{l=1}^{L} \lambda_t \phi_{l,t} \left( \tilde{x}_{l,t} - \sum_{k=1}^{K} x_{l,k,t} \right)$$

$$+ \sum_{t=1}^{T} \lambda_t \Theta_t \left( E_t - \sum_{f=1}^{F} \Lambda_{f,t} Q_{f,t} \right), \qquad \text{(D.4)}$$

where $\Theta_t \in [0,1)$ is a measure of the agent's over-assignment of energy surplus during $t$. The FOC to choose the optimal quantity of good $f$ during $t$ shows that in equilibrium marginal energy transfers of final goods equal effective energy assignment

$$\tau_{f,t} = (1 - \Theta_t) \Lambda_{f,t}. \qquad \forall\, f, t \qquad \text{(D.5)}$$

This solution ensures that, given $\boldsymbol{\Lambda_t}$, the agent produces the maximum amount of final goods while complying with energy assignment and prime mover constraints. The resolution of the Lagrangian in (D.4) involves a system of $L \cdot F \cdot T^3$ equations and unknowns. The system yields optimal production of final goods and the schedule of over-assignment of energy.